\documentclass[conference]{IEEEtran}
\IEEEoverridecommandlockouts
\usepackage{cite}
\usepackage{balance}
\usepackage{amsmath,amssymb,amsfonts}
\usepackage{algorithmic}
\usepackage{graphicx}
\usepackage{textcomp}
\usepackage{xcolor}
\def\BibTeX{{\rm B\kern-.05em{\sc i\kern-.025em b}\kern-.08em
    T\kern-.1667em\lower.7ex\hbox{E}\kern-.125emX}}

\makeatletter
\def\ps@IEEEtitlepagestyle{
  \def\@oddfoot{\mycopyrightnotice}
  \def\@evenfoot{}
}
\def\mycopyrightnotice{
  {\footnotesize
  \begin{minipage}{\textwidth}
  \centering
  Copyright~\copyright~2019 IEEE.  Personal use of this material is permitted.  Permission from IEEE must be obtained for all other uses, in any current or future media, including reprinting/republishing this material for advertising or promotional purposes, creating new collective works, for resale or redistribution to servers or lists, or reuse of any copyrighted component of this work in other works.\\
DOI: 10.1109/ETFA.2019.8869087 \\
URL: https://ieeexplore.ieee.org/document/8869087
  \end{minipage}
  }
}

\begin{document}
\sloppy

\title{A Meta-model for Process Failure Mode and Effects Analysis (PFMEA)
%
\thanks{The work leading to this paper was partially funded by the German Federal Ministry of Education and Research under grant number 01IS16043Q (CrESt).}
}
\author{\IEEEauthorblockN{Kai H\"{o}fig}
\IEEEauthorblockA{\textit{Siemens AG} \\
\textit{Corporate Technology}\\
Munich, Germany \\
kai.hoefig@siemens.com}
\and
\IEEEauthorblockN{Cornel Klein}
\IEEEauthorblockA{\textit{Siemens AG} \\
\textit{Corporate Technology}\\
Munich, Germany \\
cornel.klein@siemens.com}
\and
\IEEEauthorblockN{Stefan Rothbauer}
\IEEEauthorblockA{\textit{Siemens AG} \\
\textit{Corporate Technology}\\
Munich, Germany \\
stefan.rothbauer@siemens.com}

\and
\IEEEauthorblockN{Marc Zeller}
\IEEEauthorblockA{\textit{Siemens AG} \\
\textit{Corporate Technology}\\
Munich, Germany \\
marc.zeller@siemens.com}

\and
\IEEEauthorblockN{Marian Vorderer}
\IEEEauthorblockA{\textit{Robert Bosch GmbH} \\
\textit{Corporate Research}\\
Renningen, Germany \\
marian.vorderer@de.bosch.com}

\and
\IEEEauthorblockN{Chee Hung Koo}
\IEEEauthorblockA{\textit{Robert Bosch GmbH} \\
\textit{Corporate Research}\\
Renningen, Germany \\
cheehung.koo@de.bosch.com}

}

\maketitle

\begin{abstract}
Short product lifecycles and a high variety of products force industrial manufacturing processes to change frequently. 
Due to the manual approach of many quality analysis techniques, they can significantly slow down adaption processes of production systems or make production unprofitable. 
Therefore, automating them can be a key technology for keeping pace with market demand of the future. The methodology presented here aims at a meta-model supporting automation for PFMEA. The method differentiates product requirements, production steps and quality measures in such a way, that complex quality requirements can be addressed in any instance of a factory using a common meta-modeling language.
\end{abstract}


\section{Introduction}

Today's global competition, environmental concerns and individual customer requirements lead to reduced product lifecycles and a high variety of products. 
As a result, industrial manufacturing processes frequently change to produce new products or adjust the output to new market demand, especially in industry 4.0 production scenarios.
Classic process failure mode and effects analysis (PFMEA) is used to deliver high quality products and optimize production systems. 
But due to its often manual approach, it can significantly slow down adaption processes of production systems or make production unprofitable, especially for highly dynamic production scenarios or small lot-sizes. 

Therefore, automating PFMEA activities can be a key technology for keeping pace with market demand of the future. The methodology presented here contributes to such automation technologies by providing a meta-model for automatin a PFMEA. 




The method differentiates product requirements, production steps and quality measures in a way, that complex quality requirements can be addressed in different production scenarios using a common meta-modeling language. Using a standardized language in a later implementation seems important if production processes spread over different vendors. 

The rest of this paper is structured as follows: First, we present related work in section \ref{related}. Then, we present in section \ref{sec.metamodel} a meta-model that provides the domain specific language required to automatically conduct a PFMEA. Section \ref{casestudy} shows how the meta-model is used in an example and what the outcomes of an automated PFMEA are. Section \ref{summary} summarizes this paper and provides a perspective for future research.

\section{Related Work}\label{related}
In \cite{CPP}, the authors use FMEA among other techniques to assess the manufacturability and estimating the cost of a conceptual design in early product design phases. This manual task is used to prioritize different manufacturing options. Their work can be used in combination with the approach presented here to include costs of potential failures during manufacturing. 
In \cite{PPR}, a process resource-based approach is presented that uses an ontology to model the manufacturing capabilities and the required process steps to produce a product. Similar to the approach presented here, the authors use a standardized language set in an ontology to (semi-)automate the process of mapping production steps for a product to machinery.  Nevertheless, they do neither aim for quality aspects of the output nor for rejected items in the mapping process. 

In \cite{HRC}, another automated process is presented to identify potential hazards to operators maintainers and potential bystanders especially for a human-robot collaboration workplace. 

An approach that addresses aging effects of machinery and the effects on products can be found in \cite{afsharizand2019manufacturing}. This approach aims at modeling and evaluating the degradation of machining resources to improve process planning. This framework can be used to model failure probabilities of a PFMEA over time more precisely.

The language presented here is an approach to integrate data from design process and data from manufacturing in a standardized way to enable computer-aided process planning on a larger scale. 
In \cite{marazo2018step} the authors address the generation of machine instructions for single machinery from design documentation using digital twins of the production machinery whereas approaches like \cite{BEHANDISH2018115} aim for the identification of the right production equipment like additive or subtractive technologies to manufacture a work-piece. Many more of such approaches that focus on connecting computer-aided design (CAD) and computer-aided manufacturing (CAM) for a single machine or production equipment can be found in \cite{Al-wswasi2018}. The language as presented here also aims for a generative approach but with the focus on the larger scale of an entire value added chain. We do not aim for the automated \emph{extraction} of information from a computer-aided design, but for an automated \emph{selection} of production equipment that is capable to manufacture a product with the required qualities.

\section{Meta-model for process FMEA}\label{sec.metamodel}
The process of an automated PFMEA aims at deriving an automated selection of production machinery with the desired quality (process output) from a description of production requirements (process input) using an abstract description language (process precondition). With this automation we aim for producibility, quality and economic efficiency of a product.

\begin{figure*}[t!]
\centerline{
\includegraphics[scale=0.9]{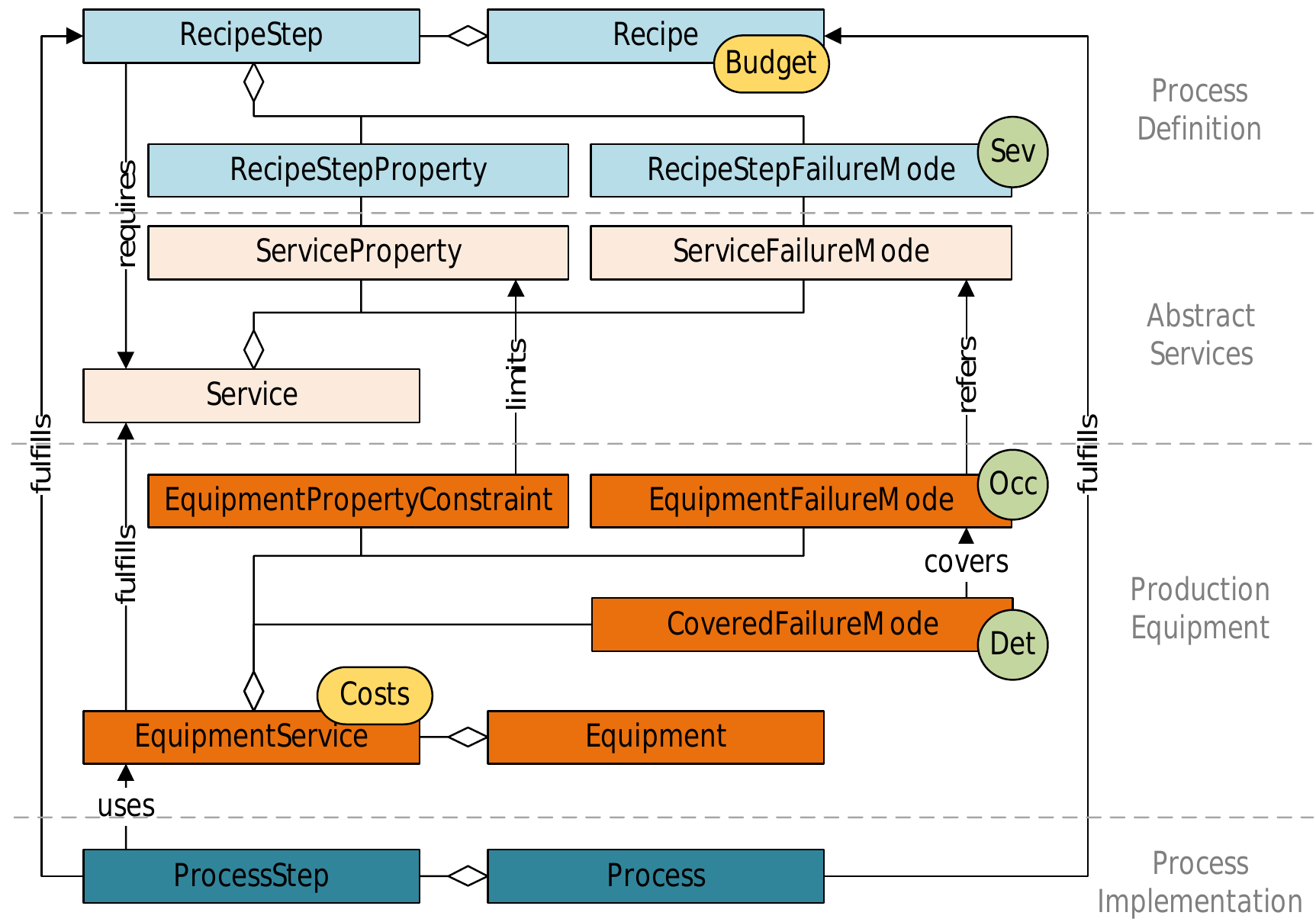}
}
\caption{Meta-model for process FMEA supporting automation}
\label{fig.pfmea}\label{metamodel}
\end{figure*}

The meta-model depicted in  figure \ref{metamodel} defines a domain specific language to link machinery with production requirements.

Abstract services provide a global library of all services. Producers that own production equipment are intended to describe the provided services in an abstract and standardized way. Each production \textit{equipment} provides different services at different costs. An \textit{equipment service} fulfills an abstract \textit{service} and concrete \textit{equipment property constraints} limit the abstract \textit{service properties}. 

For example, the abstract service pick-and-place would require parameters like weight of the item to move, a start vector, destination vector and so on. A concrete equipment fulfilling this service, e.g. a robot arm, would have limitations of these parameters like a weight below 50kg. 

Furthermore, \textit{service failure modes} describe how a \textit{service} can fail. For example, a failure mode of the service \textit{drill} could be \textit{skew drill hole}. An equipment failure mode refers to a \textit{service failure mode} and the owner of the production equipment provides information about his machinery regarding the failure scenarios, for example by providing an occurrence ($Occ$)ratio for the calculation of a risk priority number. 

Equipment services can also include quality measures, e.g., inspections, measurements, corrections or rejecting items. Such quality measures detect the occurrence of failures and cover equipment failure modes by rejecting items of minor quality. If a \textit{covered failure mode} can cover a equipment failure mode, a detection rating $Det$ is used. If a failure mode is not covered by any quality measure, the detection value is set to maximum, which represents the lowest detection rating here. Quality measures that affect production can also be applied during the product design process, e.g. when parts are designed in a way that they can only be assembled in one way (poka yoke). In this case the occurrence parameter of an erroneous assembly can be high, since it is already prevented in the design. Using this language, the owner of production equipment can model their production capabilities in an abstract way and also provide information about costs and the provided quality. 

The process definition then is done by the product development team. This team uses the abstract language interface to define the production process and the actual machinery used to perform production steps. The language is intended to be constantly extended with all new services that originate in future production technologies.

A development team or a product owner uses the language of abstract services to formulate the manufacturing process of a product using a \textit{recipe} that consists of \textit{recipe steps}. Each recipe step addresses a \textit{service} and also provides some more detailed information about \emph{what} to do in the step using \textit{recipe step properties}. The failure modes that can occur during a recipe step are extracted from the abstract definition of the referenced service. The development team can then decide for each \textit{recipe step failure mode} that belongs to a service failure mode how severe the occurrence is using a rating to later calculate a risk priority number. This first step of a risk assessment according to a process FMEA can be performed without knowledge about the concrete equipment that later manufactures a product, since it is only a requirement.

A concrete production \textit{process} then has different \textit{process steps} $P=p_1,\dots ,p_m$ that use concrete equipment services to fulfill a recipe step of a recipe $R=r_1,\dots ,r_n$. A process can \emph{produce} a recipe if the relative order of the recipe steps is the same as in the process steps (interrupted only by an arbitrary number of quality measures) and the referenced equipment services match the services and service properties of the recipe step. 

Using the severity of a failure mode from the product specification (recipe step failure mode) multiplied by the occurrence value of the equipment failure mode and the detection value a quality measure, a process FMEA with risk priority numbers can be conducted automatically for a product produced by a certain process on a concrete set of equipment. 
\[RPN = Occ * Sev * Det\]

The production will produce outputs of the desired quality, if the calculated risk priority numbers are below a certain threshold and severe failure modes occur rarely. The calculation of this risk priority number (RPN) allows to decide which machinery is capable to produce a product with the desired 
\emph{quality}. 

To improve the quality of a process, we investigate whether the risk, severity times occurrence, is smaller than $t$ (requirement) for each severity value of a recipe failure mode and occurrence value of a equipment failure mode in the process. If the risk is higher than $t$ we look for a possibility to cover that failure mode, based on the equipment available. If a suitable equipment is available that offers a quality measure covering this failure mode, then the equipment failure modes risk is reduced by an improved detection rating. In this way, quality measures that address failure modes can be added until a desired quality for the product can be reached.

Furthermore, using the estimates for \emph{costs} of machinery and quality measures, the number of rejected items by a quality measure that relates to the occurrence of an equipment failure mode and the \emph{budget} that a recipe has, the economic efficiency of a process can be calculated. 

Using the meta-model as described here, all possible processes $P_1,\dots,P_n$ are generated for a certain factory or an arbitrary set of production equipment that fulfill a recipe $R$. Combinations of quality measures can be used to extend processes that do not meet the quality requirements.
The remaining processes are ordered according to desired properties, e. g. price, quality, throughput or other resources to select an optimal process. 

In the next section, we apply our model-based approach for automated PFMEAs in a case study.
 
\section{Example}\label{casestudy}

\begin{table*}[ht]
\includegraphics[scale=0.86]{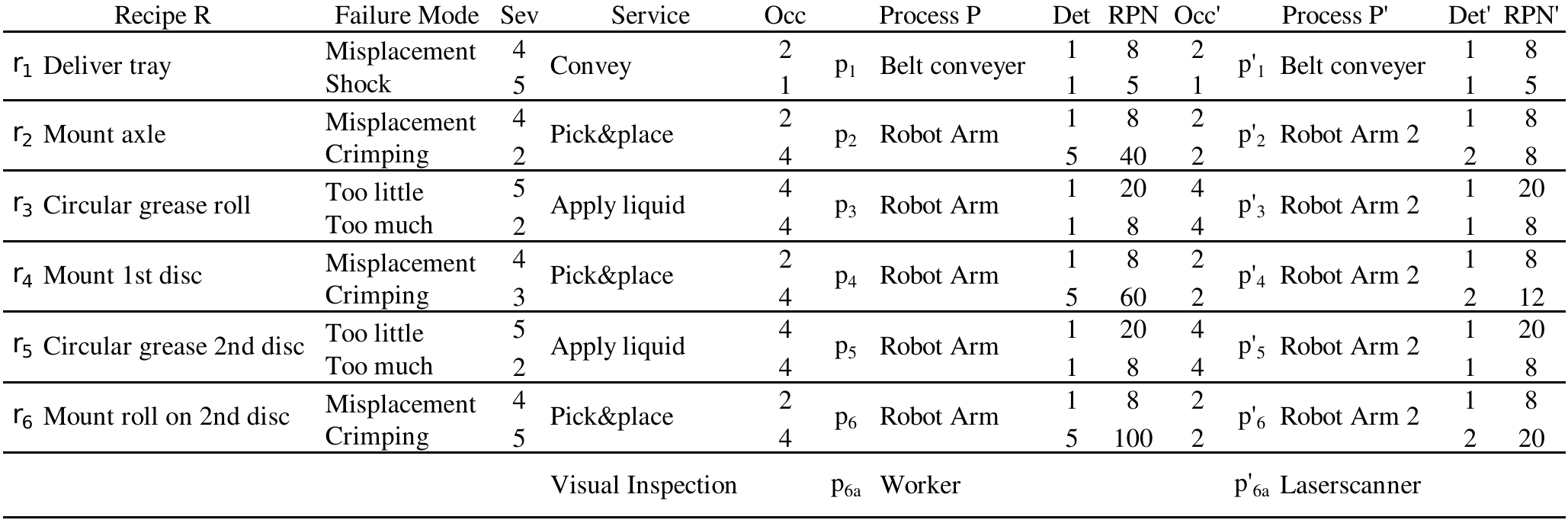}
\caption{Example product recipe and two processes using abstract services.}
\label{fig.example_neu}
\end{table*}


In this example, we want to demonstrate how the  meta-model is used to investigate the production of a small roll that consist of a roll body, an axle and two metal discs. The entire material is delivered on a tray and is set together by a robot arm that also greases the contact area of the parts. After that, a visual inspection detects insufficient products. 

The recipe steps for production are depicted on the left side in table \ref{fig.example_neu} for recipe $R=r_1,\dots,r_6$. For the first process $P=p_1,\dots,p_{6a}$ , the tray is delivered using an abstract service \emph{convey} which is implemented by the equipment \emph{belt conveyer}. The failure modes of this service are \emph{misplacement} and \emph{shock} rated by the design team with a severity value of four and five respectively. The production equipment produces failures with a occurrence of two and one. A visual inspection can safely detect both failure modes (detection=1).

The next step is to mount the axle inside the roll. This step is fulfilled by the service \emph{pick and place} which is implemented by a robot arm. Here, the object can either be misplaced or can be crimped by the clutch, which cannot be detected by a visual inspection (detection value is 5). Both discs need to be greased and there can be too much and too little grease. Having too little grease is quite severe and the worker can detect it. Having too much grease is just a minor failure. Since the roll itself is made from plastic material, crimping is severe since the roll can be damaged. This failure mode can hardly be detected (detection value is 5). 

The elements of properties and constraints are not depicted in the table for the reason of space limitations.

With the failure mode information provided by the service definition, the design team can specify \emph{what} failure mode is severe (requirement) and the vendor can specify \emph{how} often the failure mode appears on its machinery and \emph{how} the effect of the failure mode can be prevented in later products. The process $P$ generally is capable to implement the recipe $R$ since the equipment fulfills the required service of each recipe step and the relative order of the process steps matches the order of the recipe steps with an additional step at the end of the process: $p_{6a}$. 

Also depicted in table \ref{fig.example_neu} is an additional process $P'=p'_1,\dots,p'_{6a}$ that also fulfills recipe $R$ but with different equipment. A different robot arm is used, that has a lower probability of crimping. Additionally, the visual inspection is implemented by a more precise laser scanner that better detects crimping. With these two adoptions in place, the highest risk priority number is lowered from 100 to 20.

This example shows, how using a language of abstract service definitions allows to define an abstract production recipe without addressing concrete production equipment. The product design team uses abstract service definitions and properties to formulate production requirements. It can be decided (semi-)automatically if the production equipment can manufacture a product defined by a recipe. By providing information about the severity of certain failure modes, those requirements are extended by quality requirements. In a second step, a factory can map its production equipment to this abstract language and evaluate if it can produce the recipe. By providing information about the occurrence of failure modes of the existing production equipment, it can be evaluated using RPNs if the required quality can be met or if additional quality measures need to be implemented to increase the quality. By having a budget for a recipe, the vendor of a product can evaluate the economic efficiency of its possible production scenarios and decide to produce a product or to decline an offer. By comparing the RPNs of prospective processes and their economic deficiencies, an optimal process can be selected. 
 
\section{Summary}\label{summary}
In this paper, we presented a meta-model that allows to automate a design space exploration of possible production scenarios for products. Even though the design space exploration of possible processes is currently performed in a prototype and complexity is expected to be exponentially, we think it is worth continuing research since such calculations are executed during design time and not during runtime. 

Automated PFMEA allows to measure impacts of design changes to the production directly during the design phase and vendors can make new offers using the common language. This requires new data to be generated and maintained, but with the possibility for a new way of digitizing the manufacturing business and maintaining quality requirements.

\balance
\bibliographystyle{IEEEtran}
\bibliography{Bibliography,safety}

\end{document}